\documentclass[fleqn,twoside]{article}
\usepackage[headings]{espcrc2}

\readRCS
$Id: espcrc2.tex,v 1.2 2004/02/24 11:22:11 spepping Exp $
\usepackage{graphicx}
\usepackage[figuresright]{rotating}

\newcommand{\AmS}{{\protect\the\textfont2
  A\kern-.1667em\lower.5ex\hbox{M}\kern-.125emS}}

\title{Precise predictions for LHC using a {\tt GOLEM}}

\author{\underline{T. Binoth}\address[EDI]{School of Physics, The University of Edinburgh, 
        Edinburgh EH9 3JZ, UK}%
        \thanks{This work was supported by the British Science and Technology Facilities Council (STFC) and
        the Scottish Universities Physics Alliance (SUPA).},
        A. Guffanti\address[FREI]{Department of Physics, University of Freiburg, 
        Hermann-Herder-Str. 3a, D-79104 Freiburg, Germany},
        J.-Ph. Guillet\address[LAPTH]{LAPTH, 9, Chemin de Bellevue BP 110, 74941 Annecy le Vieux, France},
        G.~Heinrich\address[IPPP]{IPPP, University of Durham, Durham DH1 3LE, UK},
   	S.~Karg\address{Institute for Theoretical Physics E, RWTH Aachen, D-52056 Aachen, Germany},
 	N.~Kauer\address{Institute for Theoretical Physics, University of W\"urzburg, 
	D-97074 W\"urzburg, Germany},
	P.~Mertsch\address{Rudolf Peierls Centre for Theoretical Physics,
	University of Oxford, 1Keble Road, Oxford, OX1 3NP, UK},
        T.~Reiter\addressmark[EDI],
	J.~Reuter\addressmark[FREI],
	and
	G.~Sanguinetti\addressmark[LAPTH]}
       
\runtitle{Precise predictions for LHC using a {\tt GOLEM}}
\runauthor{T. Binoth}

\begin{document}

\begin{abstract}
\begin{flushright}
Edinburgh 2008/24, IPPP/08/49, DCPT/08/98, LAPTH/CONF-1260/08, PITHA 08/17
\end{flushright}
In this talk we present recent next-to-leading order results relevant
for LHC phenomenology obtained with the {\tt GOLEM} method. After 
reviewing the status of this Feynman diagrammatic approach for multi-leg
one-loop calculations we discuss three applications: the loop-induced process $gg\to Z^*Z^*$
and the virtual corrections to the five and six point processes
$qq \to ZZg$ and  $u\bar{u} \to s\bar{s} c\bar{c}$.
We demonstrate that our method leads to representations of
such amplitudes which allow for efficient phase space 
integration. In this context we propose a reweighting technique of the
leading order unweighted events by local K-factors. 

\vspace{1pc}
\end{abstract}

\maketitle

\section{Introduction}

The Large Hadron Collider will explore our understanding of
fundamental interactions in the multi-TeV range. Apart from testing
the Higgs mechanism which is the final cornerstone of the Standard Model (SM),
also a plethora of extensions of the SM will be put under scrutiny.
Whatever final states will be detected, the initial state will 
always consist of QCD partons. The perturbative description of such
processes is necessarily plagued by renormalisation and factorisation
scale uncertainties. Only if next-to-leading order (NLO) corrections are included
the logarithmic dependence on these scales is tamed and one 
arrives at sufficiently reliable predictions for various signal and
background processes. For a discussion of what remains to be done see \cite{Bern:2008ef}.  

For a full next-to-leading order evaluation of an N-point process one has to 
combine virtual corrections with real-emission corrections using some
infrared subtraction method. The tree-like processes can be 
evaluated by using standard leading order tools. Meanwhile, 
automated ways to deal with the IR subtractions based on the Catani-Seymour dipole approach \cite{Catani:1996vz} 
are also on the market \cite{Gleisberg:2007md,Seymour:2008mu}.  
The evaluation of the one-loop contribution for N-point processes is not yet at this level
of automation, although new ideas emerged and a lot of progress has been made recently in various
directions 
\cite{Berger:2008sj,Catani:2008xa,Giele:2008bc,Ellis:2008ir,Britto:2008vq,Ossola:2006us,Denner:2005nn,Bern:2007dw}. 

\section{The {\tt GOLEM} method}

The aim of our collaboration is to provide a tool which allows for a numerically
stable evaluation of multi-leg one-loop amplitudes: {\tt GOLEM}\footnote{{\tt GOLEM}=General 
One-Loop Evaluator for Matrix elements.}. It is based on the method described in \cite{Binoth:2005ff,Binoth:1999sp}.
The approach relies on the evaluation of Feynman diagrams and the reduction of tensor
integrals using a form factor approach. The form factors can be evaluated in various ways
as outlined below.

We organize the evaluation of a one-loop amplitude as follows:
\begin{itemize}
\item generate Feynman diagrams using {\tt QGRAF} \cite{Nogueira:1991ex} or {\tt FeynArts 3.2} \cite{Hahn:1998yk}.
\item separate and perform colour algebra
\item project on helicity amplitudes 
\end{itemize}
At this point two independent set-ups exist.
Firstly,
a completely symbolic reduction to standard scalar integrals with up to four external legs 
can be obtained using {\tt FORM} \cite{Vermaseren:2000nd}
and {\tt MAPLE}.
\begin{eqnarray}
\mathcal{M}^{\{\lambda\}} &\to& C_{box} I_4^{n+2} 
                                                      + C_{tri} I_3^{n} \nonumber \\&&
                                                      + C_{bub} I_2^{n} 
                                                      + C_{tad} I_1^{n} + \cal{R}\; .\nonumber
\nonumber\end{eqnarray}
The respective coefficients are rational polynomials in Mandelstam variables. The extraction 
of the rational part, $\cal{R}$, of the amplitude can be done separately \cite{Binoth:2006hk}. 
As long as no efficient  tools for the  manipulation of multivariate rational polynomials 
are available, interactive user input is needed to produce sufficiently compact amplitude 
expressions in the purely symbolic approach.   
Secondly, apart from the symbolic approach, we provide a numerical tensor reduction.
Schematically the amplitude is expressed in terms of form factors which 
resemble Feynman parameter integrals with Feynman parameters in the numerator.
\begin{eqnarray}
\mathcal{M}^{\{\lambda\}} &\to& C_{box}^{ijk} I_4^{n+2,n+4}(x_ix_jx_k) \nonumber\\ &&
                                                      + C_{tri}^{ijk} I_3^{n,n+2}(x_ix_jx_k) + \dots
\nonumber\end{eqnarray}
These form factors are implemented in a 
{\tt FORTAN90} code and can be evaluated by numerical reduction and also by using 
one-dimensional integral representations. The form factors were designed to avoid
the occurrence of so-called Gram determinants which usually hamper a numerically stable
evaluation of large Feynman diagrammatic expressions. 
 Our method has been successfully applied to various calculations with up to
six point functions \cite{Binoth:2006ym,Andersen:2007mp,Binoth:2007ca,Bernicot:2007hs}. 

\section{Applications for LHC phenomenology}

We discuss now recent evaluations of three loop amplitudes which are 
relevant in the context of Higgs searches at the LHC. 

\subsection{The process $gg \to Z^*Z^*$}
In \cite{Binoth:2006mf,Binoth:2005ua,Duhrssen:2005bz} it has been shown that the gluon induced production 
of charged vector boson pairs accounts  for about 30 percent of the background 
to Higgs searches in that channel after cuts. Although a similar calculation
for neutral vector bosons has been performed a long time ago \cite{Zecher:1994kb}, 
no public code is available which motivated us to redo this calculation using our method. 

As only 4-point functions are present, a symbolic expression for the amplitude
could be obtained, where numerically dangerous denominators have been cancelled
algebraically. 
The expressions are implemented in a flexible computer program
{\tt GG2ZZ} \cite{gg2zz_link} which also contains the photonic contributions.
The size of the gluon contribution to the $ZZ$ cross section in relation
to the quark induced part is included in the following table:

\smallskip

\begin{center}
\renewcommand{\arraystretch}{1.4}
\begin{tabular}{|c|c|cc|c|c|}
 \cline{2-6}
\multicolumn{1}{c|}{} & \multicolumn{5}{c|}{$\sigma(pp \to Z^\ast(\gamma^\ast)Z^\ast(\gamma^\ast) 
\to \ell\bar{\ell}\ell'\bar{\ell'})$~[fb]} \\ \cline{2-6}
\multicolumn{1}{c|}{} & & 
\multicolumn{2}{c|}{\raisebox{1ex}[-1ex]{$q\bar{q}$}}
& \multicolumn{1}{c|}{} &\multicolumn{1}{c|}{} \\[-1.5ex]
\cline{3-4}
\multicolumn{1}{c|}{} & 
\multicolumn{1}{c|}{\raisebox{2.7ex}[-2ex]{$gg$}} & 
\raisebox{0.9ex}{LO} & \raisebox{0.9ex}{NLO} 
& \raisebox{2.25ex}[-2ex]{$\frac{\sigma_{\rm NLO}}{\sigma_{\rm LO}}$} & 
  \raisebox{2.25ex}[-2ex]{$\frac{
 \sigma_{{\rm NLO}+gg}}{\sigma_{\rm NLO}}$}
\\[-1.5ex]
  \hline
 $\sigma_{\rm std}$ & 1.49 & 7.34 & 10.95 & 1.49 & 1.14 \\
 \hline
\end{tabular}
\end{center}

\smallskip

For the numerical results we use the following set of input parameters:
$M_W = 80.419$ GeV, $M_Z = 91.188$ GeV, $G_F = 1.16639\,10^{-5}$ GeV$^{-2}$, $\Gamma_Z = 2.44$
GeV. The electromagnetic coupling is defined in the $G_{\mu}$ scheme.
The pp cross sections are calculated at $\sqrt{s} = 14$ TeV employing the CTEQ6L1
and CTEQ6M parton distribution functions at tree- and loop-level, 
for more details see \cite{:2008uu}.
Applying standard cuts: $p_{T\ell} > 20$ GeV,75 GeV$< M_{\ell^+\ell^-} <$ 105  GeV, $ |\eta_\ell| < 2.5,\ \ 75$ GeV,  we find that the gluon contribution
accounts for 14\% to the total $pp \to ZZ$ process. The $q\bar{q}$ contribution was 
evaluated using {\tt MCFM} \cite{Campbell:2000bg}. 

\begin{figure}[htb]
\includegraphics[height=7.0cm, clip=true, angle=90]{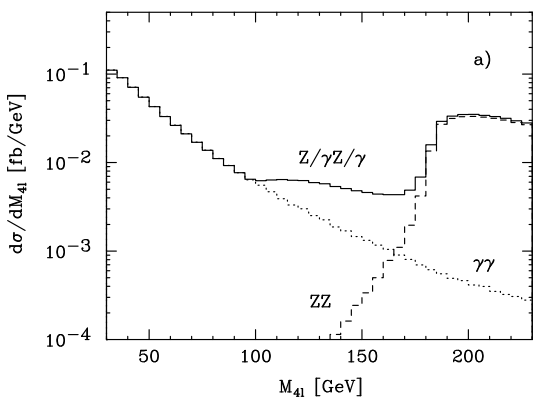}
\caption{The invariant mass distribution of the 4 leptons including photonic
contributions. Only a minimal cut $M_{\ell^+\ell^-}>5$ GeV is applied.}
\label{fig:gg2zz}
\end{figure}

The effect of the photon contribution can be seen best in
the invariant mass distribution of the 4 leptons in Fig~\ref{fig:gg2zz}. 
Between the one- and two-Z threshold the interference effects are sizable.

\subsection{The process $PP \to ZZ+$jet}

This process is of relevance in the context of Higgs searches in the Higgs plus jet channel. 
We have obtained analytic expressions for all  36 partonic one-loop helicity amplitudes $q \bar{q} Z Z g \to 0$
which contribute to this process. 
The colour structure is simple, one finds three different colour structures.
By applying projection operators to each Feynman diagram, reducible scalar products between
the loop momentum and external momenta can be expressed by inverse propagators and cancelled.
In this way only rank one five-point functions remain, together with two-, three- and four-point tensor integrals. 
We use the 'tHooft-Veltman scheme which needs an accompanying  prescription for
$\gamma_5$. By splitting the $\gamma$-algebra and the loop momentum in a 4- and (n-4)-dimensional 
part we have to add a finite counterterm $\sim (1-\alpha_s C_F/\pi)$ to the axial coupling to 
guarantee that the axial and vector part of the vector boson renormalise in the same way \cite{Larin:1993tq}.  
After UV renormalization, only IR poles remain. A finite expression can be obtained by adding the 
Catani-Seymour insertion operator, $I(\epsilon)$ \cite{Catani:1996vz}, to the result.   
We have integrated the resulting expression for the LHC energy over the phase space using
the cut $p_{T jet}>100$ GeV and find:
\begin{eqnarray}
\sigma_{LO} &=& 1003.1 \pm 0.4\; \textrm{fb} \nonumber\\
\sigma_{LO+virt.} &=& \; \; 899.0 \pm 4.7 \;\textrm{fb} \nonumber
\end{eqnarray}
Here $M_W=80.403$ GeV, $M_Z=91.1876$ GeV, $G_F=1.16637\cdot 10^{-5}$ GeV$^{-2}$ are used.
$\alpha$ is evaluated in the $G_\mu$ scheme and we used $\mu_R=\mu_F=M_Z$. For the LO result we use the {\tt CTEQ6L1}
and for NLO the {\tt CTEQ6M} parton distribution functions. $\alpha_s$ is evaluated using
the {\tt LHAPDF} routines. Note that the number does not include the closed fermion loop contribution
to this process as it turned out to be numerically irrelevant.

\begin{figure}[htb]
\includegraphics[height=5.0cm, clip=true]{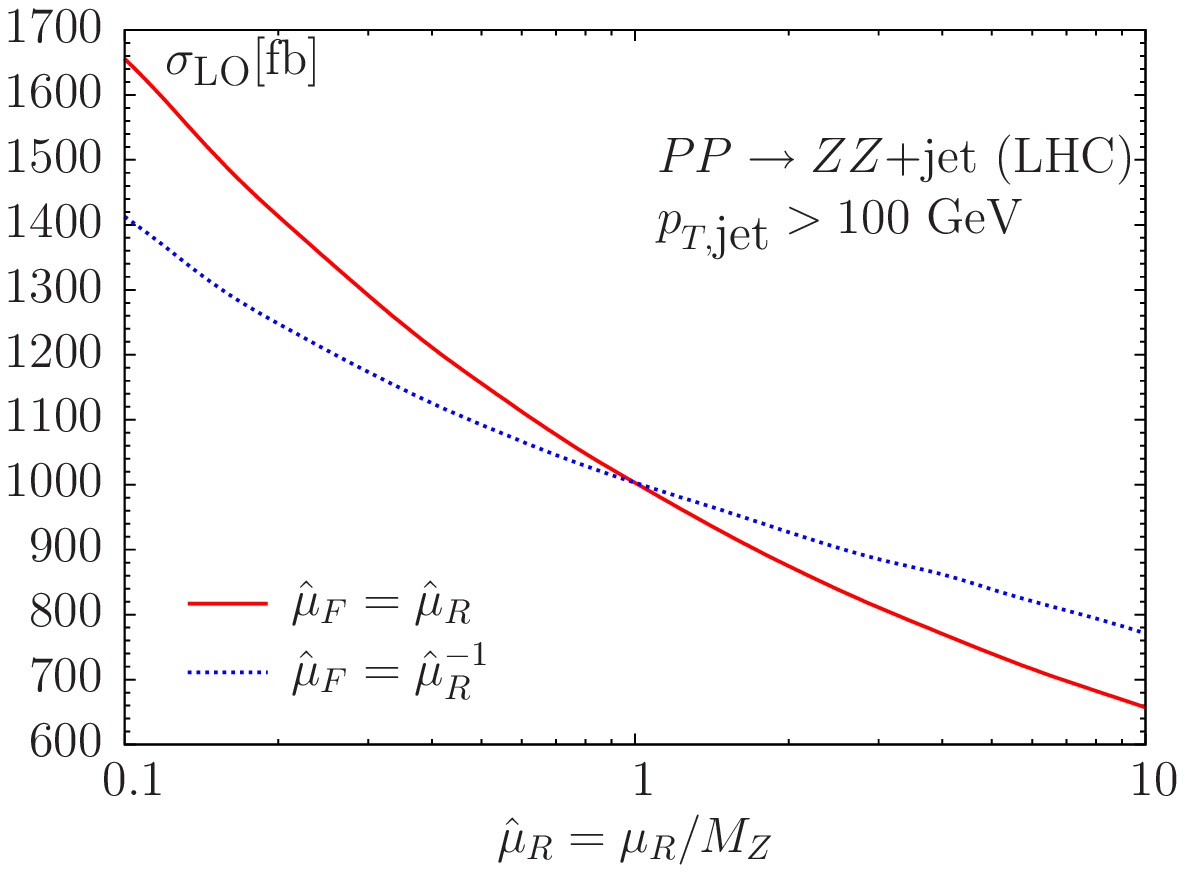}
\includegraphics[height=5.0cm, clip=true]{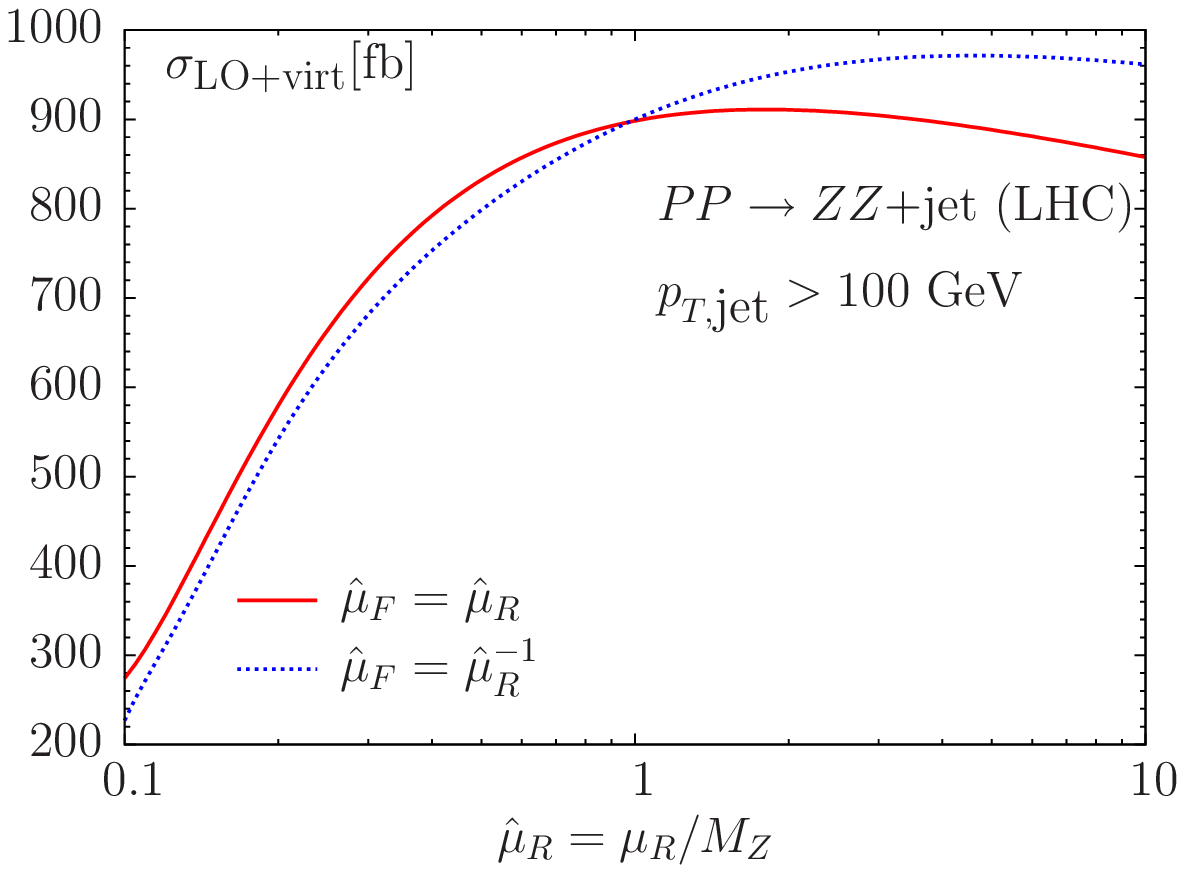}
\caption{Scale dependence of the LO and  virtual next-to-leading order corrections.
The two curves in each plot show diagonal and anti-diagonal variation
of the scales.}
\label{fig:mu}
\end{figure}

We compare the scale dependence of the LO term and
the NLO virtual part in Fig~\ref{fig:mu}.
The figures indicate that the inclusion of the virtual corrections indeed
stabilise the prediction for scales around $2 M_Z$. 
Note that the real emission part of the NLO correction will add another
$\mu_F$ dependent term which stems from the initial state singularities.
As new colour channels are present at NLO one expects 
actually a deterioration of the scale dependence, as has been observed
in the $PP\to W^+W^-+$jet  case \cite{Dittmaier:2007th}.  

Results for this related process have been presented by
two other groups already \cite{Dittmaier:2007th,Campbell:2007ev}. We have also evaluated this process 
and compared our evaluation  with both groups for single phase space points. 
Perfect agreement was found \cite{Bern:2008ef}.    

\subsection{The process $u\bar{u}\to s\bar{s} b\bar{b}$}

Signatures of beyond SM processes
contain typically leptons, jets and missing energy and one easily reaches 
large numbers of final state partons at the LHC. Up to now no complete 6-point process,
which is related to a $2\to 4$ kinematics, is 
evaluated at next-to-leading order in $\alpha_s$. Note that
progress in that direction  has been 
reported  for the process $pp\to b\bar{b} t\bar{t}$ at this workshop \cite{Bredenstein:2008ia,Bredenstein:2008zb}.
An example for another relevant 6-point process emerges in two Higgs
doublet models, which lead in certain parameter regions predominantly to 4 b-jets in the final state.
To understand the related background in detail, the process $PP\to b\bar{b}b\bar{b}$ 
has to be known at NLO. Note that at the LHC one can safely neglect the bottom mass, if
realistic $p_T$ and $b$ separation cuts are applied. The partonic amplitudes 
$u\bar{u}\to s\bar{s} b\bar{b}$ and $gg\to s\bar{s} b\bar{b}$ are sufficient 
to predict this cross section in the massless limit. We report here on the
successful evaluation of the virtual part of the first of these two amplitudes.

In detail, the $u\bar{u}\to s\bar{s} b\bar{b}$ amplitude can be written in terms 
of six colour structures. Two independent helicity
amplitudes, ${\cal A}^{++++++}$ and ${\cal A}^{++++--}$, are needed. They
were evaluated in two completely independent ways.
In one evaluation a fully symbolic reduction to scalar integrals was performed, in the other one
each Feynman diagram was mapped to a form factor representation and translated into a {\tt FORTRAN90}
code. Both calculations are highly automated such that the evaluation of other processes
only needs the respective Feynman diagrammatic input and the specification of colour and helicity 
projections. In the given case one has to evaluate 25 pentagon and 8 hexagon diagrams.
After UV renormalization and adding the IR  insertion operator all poles in $1/(n-4)$
cancel and one is left with a finite expression which can be evaluated numerically.
The {\tt FORTRAN90} code was organized such that the reevaluation of algebraic
terms was avoided by recursive organization of the expressions and caching.
The evaluation time for one phase space point of the full amplitude, summed over
helicities and colour is about 0.8 seconds on a 3.2 GHz Intel Pentium 4 processor. As the 
integration over phase space can be trivially parallelised this is sufficiently fast  
in what concerns the evaluation of distributions. 

As the evaluation of the amplitude needs a large number of numerical operations,
one typically observes numerical problems in parts of the phase space where
denominators become small and form factors, respectively scalar integrals,
are not linearly independent anymore. If one integrates directly the 
LO plus finite virtual corrections over the phase space, adaptive numerical integrators
tend to sample phase space points in these critical phase space regions. This happens   
if the induced variations influence the result at the order of the accuracy goal.
To avoid this kind of destabilisation we have applied the following method
for integrating the virtual NLO corrections.

We first evaluate the LO contribution over the target phase space
\begin{eqnarray}
\sigma_{LO} = \int d\vec{x} f_0(\vec{x})\nonumber
\end{eqnarray}
and generate unweighted events $E_{j=1,\dots,N}$.
The latter are related to a parameter transformation $\vec{x} \to \vec{y}$
on phase space such that the new variables have constant density 
$d\vec{y}\sim d\vec{x} f_0(\vec{x})/\sigma_{LO}$. Any observable $\mathcal{O}$ 
can be estimated by distributing the unweighted events, $E_j$, into the respective bins.
\begin{eqnarray}
\langle \mathcal{O}\rangle_{LO} = \frac{\sigma_{LO}}{N} \sum\limits_{j=1}^N \chi(E_j),\,
\chi(E_j)=\left\{ \begin{array}{l} 1,\, E_j \in \mathcal{O} \\ 0,\, \textrm{else}  \end{array} \right. \nonumber
\end{eqnarray}

An estimate of the LO plus virtual corrections can now be obtained using the 
same set of events. The relation
\begin{eqnarray}
\sigma_{LO+virtual} = \int d\vec{x} f_1(\vec{x}) = \sigma_{LO} \int d\vec{y}  K(\vec{y})\,,\nonumber
\end{eqnarray}
where $K=f_1/f_0$ is a local K-factor, implies
\begin{eqnarray}
\langle\mathcal{O}\rangle_{LO+virt.}  = \frac{\sigma_{LO}}{N} 
\sum\limits_{j=1}^N \chi(E_j) K(E_j)\nonumber ,
\end{eqnarray}
which is a simple reweighting of a LO event sample. For this purpose the LO events should be evaluated with NLO pdfs.  
In this way no integration over the finite virtual corrections is needed, one simply
has to evaluate the virtual corrections for each unweighted event which 
belongs to a specified observable. Of course it still can happen that
this evaluation is plagued by numerical problems but it does not negatively affect
the sampling of test points in integration methods. This method leads to a good estimate if the virtual corrections are
sufficiently close to the LO distribution, such that the unweighted events 
are also representative for the LO+virtual differential cross section. Note that this has to be 
fulfilled for any observable for which perturbation theory is meaningful in the first place.    

\begin{figure}[htb]
\includegraphics[height=5.0cm, clip=true]{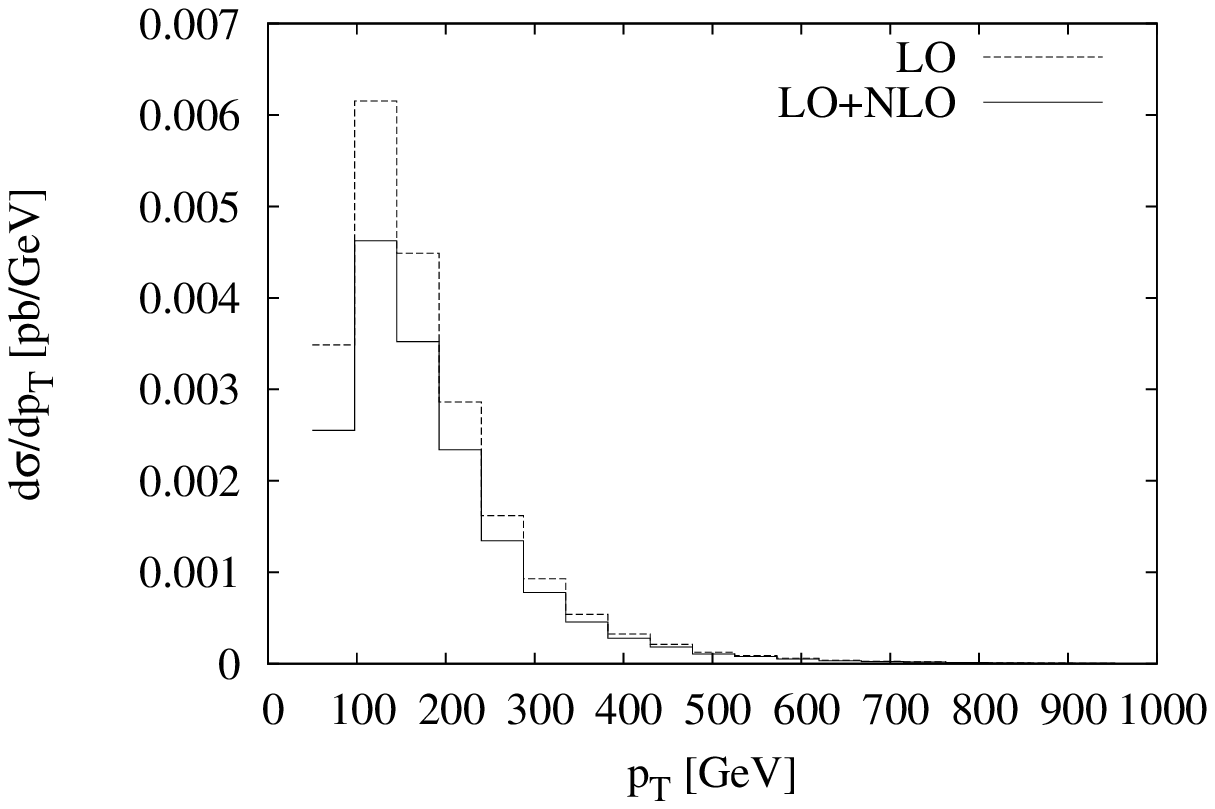}
\includegraphics[height=5.0cm, clip=true]{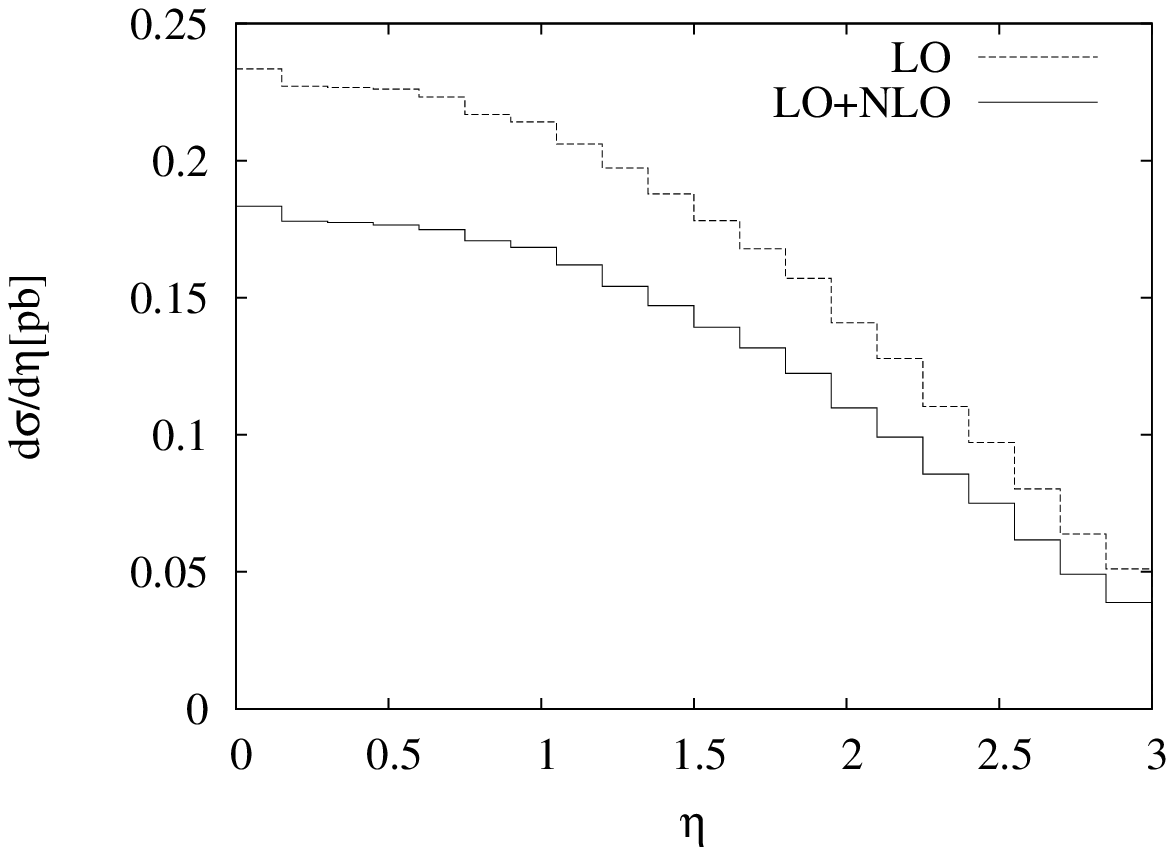}
\caption{The $p_T$ and rapidity distribution of the leading jet. The full line
is the LO, the dashed line is obtained by adding the UV/IR finite contribution of the virtual
part of the NLO prediction, as defined in the text.}
\label{fig:6q}
\end{figure}

To illustrate this method we show 
in Fig.~\ref{fig:6q} the effect of the virtual contribution on the distribution
of the leading jet, i.e. the jet with the highest energy.
To define the LO 4-jet observable we use  the cuts: $p_{Tj}>50$ GeV, $|\eta_j| < 3$
and $\Delta R=\sqrt{\Delta\phi^2 + \Delta\eta^2}>0.3$. The LO cross section and the corresponding
unweighted events were evaluated
using {\tt WHIZARD} \cite{Kilian:2007gr} with {\tt CTEQ6M} pdfs and the scale choice
$\sum_{j=1}^4 p_{Tj}/4$, $\mu=100$ GeV and one-loop running for $\alpha_s$. 
The LO + finite virtual contribution was evaluated as
described above. For the LO and LO+virtual contribution we obtain
\begin{eqnarray}
 \sigma_{LO} &=& 88.5\pm 0.2 \, \textrm{fb} \nonumber \\
\sigma_{LO+virtual} &=&  69.0\pm 0.2 
 \, \textrm{fb} \nonumber
\end{eqnarray}
The histograms in Fig.~\ref{fig:6q} are filled with 200,000 unweighted events. When evaluating the local K-factors,
less than 1\% of all points showed an indication of numerical instability.
These critical points where simply reevaluated by using the quadruple precision version
of our code. A similar evaluation of the $gg\to s\bar{s} b\bar{b}$ amplitude 
and the combination with the real emission corrections is in progress.

\section{Conclusion}

In this talk we have presented recent results of the {\tt GOLEM} collaboration.
The implementation of our method to evaluate Feynman diagrammatic representations
of amplitudes in symbolic/numerical computer programs has
been completed in the context of one-loop amplitude evaluations relevant for the LHC.
Here we presented results for the process $gg\to Z^*Z^*$, and the virtual 
corrections to $PP\to ZZj$ and $u\bar{u}\to s\bar{s} b\bar{b}$. 
We have proposed a new indirect integration method for virtual corrections which 
is based on the evaluation of local K-factors for unweighted events defined by the LO 
cross section.  We conclude that our method is numerically efficient  and can  provide
predictions for multi-leg one loop processes at TeV colliders.

\end{document}